\documentclass[a4paper,showpacs,twocolumn,amsmath,amssymb,floatfix,prb,preprintnumbers,footinbib]{revtex4}
\usepackage{graphicx}
\usepackage{amsmath,amsfonts}
\usepackage{color}

\newcommand{\e}{\varepsilon}
\newcommand{\m}[1]{\mathrm{#1}}
\newcommand{\ket}[1]{\left|#1\right\rangle}
\newcommand{\bra}[1]{\langle #1|}

\newcommand{\bs}[1]{\boldsymbol{#1}}

\begin{document}

\title{AC Josephson transport through interacting quantum dots}

\author{Bastian Hiltscher$^1$, Michele Governale$^2$, and J\"urgen K\"onig$^1$} 
\affiliation{$^1$Theoretische Physik, Universit\"at Duisburg-Essen and CENIDE, D-47048 Duisburg, Germany\\
$^2$School of Chemical and Physical Sciences  and MacDiarmid Institute for Advanced Materials and Nanotechnology, Victoria University of Wellington, P.O. Box 600, Wellington 6140, New Zealand}

\date{\today}
\begin{abstract}
We investigate the {\em AC} Josephson current through a quantum dot with strong Coulomb interaction attached to two superconducting and one normal lead.
To this end, we perform a perturbation expansion in the tunneling couplings within a diagrammatic real-time technique.
The {\em AC} Josephson current is connected to the reduced density matrix elements that describe superconducting correlations induced on the quantum dot via proximity effect.
We analyze the dependence of the {\em AC} signal on the level position of the quantum dot, the charging energy, and the applied bias voltages.  
\end{abstract}

\maketitle
\section{Introduction}
Josephson junctions can be formed by linking two superconductors via an insulator, a normal conductor or a constriction in an otherwise continuous superconducting material.\cite{josephson62,tinkham96} 
Advancements in nanofabrication enabled to contact superconductors with quantum dots (QDs), which can be formed in carbon nanotubes,\cite{buitelaar02,pillet10,cleuziou06,eichler07,herrmann10,jorgensen07} in InAs nanowires,\cite{sand07,hofstetter09,vandam06,heiblum12} in graphene,\cite{dirks11} or by means of self-organization in InAs with Al electrodes.\cite{buizert07,deacon10,kanai10}
One motivation to investigate hybrid superconductor-QD devices\cite{franceschi10,martin11} is the possibility to tune their properties via external electrodes that shift the discrete energy levels of the QD.
Another feature characteristic for QDs is the charging energy that may give rise to effects based on the interplay of  Coulomb repulsion and superconducting correlations.
It is, therefore, an interesting question to ask how the discrete level spectrum and the charging energy affect the {\em DC} and {\em AC} Josephson transport between two superconductors coupled via a QD.

In the absence of a bias voltage, a finite {\em DC} current can be sustained in such a S-QD-S by the {\em DC} Josephson effect.
This has been confirmed experimentally,\cite{vandam06,jorgensen07} which shows that two electrons forming a Cooper pair can tunnel coherently one by one through a strongly interacting quantum dot.
When neglecting the charging energy, the {\em DC} Josephson effect in such a S-QD-S system can be studied within a scattering approach.\cite{beenakker92} 
But also Coulomb-interaction effects have been included in various formalisms as perturbation expansions in the tunneling Hamiltonian\cite{glazman89,rozhkov01,pala07,governale08} and in the Coulomb repulsion,\cite{matsumoto01,vecino03} a mean-field approach,\cite{rozhkov99} quantum Monte-Carlo simulations,\cite{siano04} a renormalization-group technique,\cite{sellier05} or numerical diagonalization of an effective dot Hamiltonian.\cite{tanaka07,karrasch08,meng09} 

A finite bias voltage gives rise to a more complicated transport behavior. 
In addition to a finite {\em DC} current, sustained by quasiparticle tunneling and (multiple) Andreev reflection, there is a time-dependent component due to the {\em AC} Josephson effect.
Theoretical works for this regime have mainly concentrated on the limit of vanishing or weak Coulomb repulsion.
The DC component has been studied by focusing on single quasiparticle tunneling,\cite{whan96,levy97,kang98} using a slave boson mean-field approximation, \cite{eichler07,avishai03} or performing a perturbation expansion in the charging energy.\cite{dell08}
Multiple Andreev reflection processes not only give rise to a stationary current but also lead to higher harmonics contributing to the {\em AC} Josephson transport. 
A quantitative description including this interplay has been investigated in quantum point contacts by means of a scattering\cite{bratus95,averin95} or a Hamiltonian approach.\cite{cuevas96}
In a noninteracting quantum dot the dependence of the different harmonics on the bias voltage has been studied.\cite{sun02} 
Also dephasing effects introduced by a third, normal electrode added to the S-QD-S setup have been investigated within a Keldysh formalism applied to a noninteracting system to find that for gradually increasing coupling to the normal conductor the {\em AC} signal decreases.\cite{jonckheere09}
Further works deal with the time evolution of the current after switching on a finite bias voltage,\cite{perfetto09,stefanucci10} or with polaronic effects due to coupling to vibrational modes.\cite{wu12}
 
\begin{figure}
\
\includegraphics[angle=270,width=.95\columnwidth]{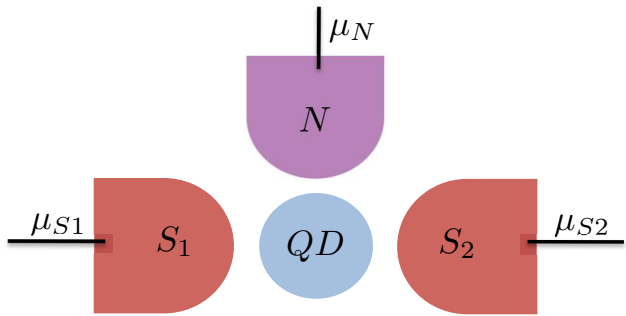}
\caption{
\label{additionalN}
(Color online) A sketch of the device under investigation: a quantum dot with strong Coulomb interaction is weakly tunnel coupled to one normal and  two superconducting leads.}
\end{figure}

The aim of this paper is to analyze the {\em AC} Josephson effect through QDs with strong Coulomb repulsion that cannot be neglected or treated perturbatively.
To this end, we extend a real-time diagrammatic approach for the {\em DC} current presented in Refs.~\onlinecite{pala07,governale08} to {\em AC} transport. 
We apply this formalism to a  three-terminal geometry consisting of a strongly interacting quantum dot, which is weakly tunnel coupled to two superconductors and one normal conductor, see Fig.~\ref{additionalN}, and perform a perturbation expansion in tunnel couplings to lowest order. 
To sustain ({\em DC} and {\em AC}) Josephson currents through the QD to first order in the tunnel-coupling strengths, superconducting correlations must be induced on the QD by the proximity of the superconducting leads.
We find that a large {\em AC} current between QD and one superconducting lead requires a large proximization of the QD by the other superconducting lead which can be achieved by tuning the gate voltage accordingly. 
We discuss the amplitude of the {\em AC} components of the current between the superconductors as a function of gate and bias voltage. 
 
\section{Model}
The considered system consists of a quantum dot tunnel coupled to one normal and two superconducting leads. 
Its total Hamiltonian is given by $H=H_\text{dot}+H_\text{tun}+\sum\limits_r H_r$.  The index $r$ refers to the leads and can take the values $S1,S2,N$.
The quantum dot is assumed to accommodate one spin-degenetate level $\e$. It is described by the Anderson impurity model,
\begin{equation}
H_\text{dot}=\varepsilon\sum\limits_{\sigma} d^\dagger_\sigma d_\sigma+Un_\uparrow n_\downarrow \, ,
\end{equation}
where $n_\sigma=d^\dagger_\sigma d_\sigma$ is the number operator and $d^{(\dagger)}_\sigma$ annihilates (creates) an electron with spin $\sigma$ on the dot. 
Coulomb interacting is accounted for by the charging energy $U$ for double occupation.
The Hilbert space of the isolated dot is four dimensional and is spanned by the kets $\{|\chi\rangle\}$ with $\chi=0,\uparrow, \downarrow,d $, corresponding  respectively to empty, singly occupied with spin up, singly occupied with spin down, and doubly occupied dot.

The two superconducting leads are modeled by the mean-field BCS Hamiltonian 
\begin{align}
\label{mfBCS}
H_{r}^\text{BCS}=\sum\limits_{k,\sigma}\varepsilon_{k} c_{r{k}\sigma}^\dagger c_{r{k}\sigma}- \left( \Delta_r^*\sum\limits_k S_r^\dagger c_{r-k\downarrow}c_{rk\uparrow}+ h.c. \right)\, .
\end{align}
Here, $c_{rk\sigma}^\dagger$ is the creation operator of an electron in lead $r$ with momentum $k$ and spin $\sigma$, and $\Delta_r$  is the superconducting pair potential. 
The operator $S_r^{(\dagger)}$ annihilates (creates) a Cooper pair in lead $r$, which ensures particle conservation.
By making use of a Bogoliubov transformation, the Hamiltonian can be diagonalized,
\begin{equation}
H_r=\sum\limits_{rk\sigma}E_{rk}\gamma_{rk\sigma}^\dagger\gamma_{rk\sigma}+\mu_rN_r
\end{equation}
with the quasiparticle operators $\gamma_{rk\sigma}$ and the corresponding eigenenergies $E_{rk}=\sqrt{(\e_k-\mu_r)^2+|\Delta_r|^2}$. The chemical potential of superconductor $r$ is given by $\mu_r$ and $N_r$ is the total number of electrons, which is the number of quasiparticles plus twice the number of Cooper pairs. 
Also the normal conductor can be described by Eq. (\ref{mfBCS}) by simply setting $\Delta_N=0$.

Tunneling between dot and leads is described  by the tunneling Hamiltonian
\begin{equation}
H_\text{tun}=\sum\limits_{rk\sigma}t_rc^\dagger_{rk\sigma}d_\sigma+H.c. \, .
\end{equation}
The tunneling amplitude $t_r$ as well as the density of states $\rho_r$ are assumed to be energy independent in the window relevant for transport. Furthermore, we define $\Gamma_r=2\pi |t_r|^2\rho_r$. 
Finally, we set $\hbar=1$ and we reinstate it in the units used for the figures. 

Due to tunneling, the superconducting leads can induce superconducting correlations on the quantum dot.~\cite{pala07,governale08} 
The resonance condition for this proximity effect due to superconductor $r$ is that the energy for the doubly occupied dot, $2\varepsilon+U$, equals the energy for an empty dot plus an extra Cooper pair in the condensate of the superconductor, $2\mu_r$.
It is, therefore, convenient to introduce the detunings $\delta_r =  2\varepsilon+U - 2\mu_r$ with $r=S1,S2$.

\section{Method}
\label{method}

A real-time diagrammatic approach to {\em DC} transport through quantum dots tunnel coupled to normal and superconducting leads, has been introduced in Ref.~\onlinecite{governale08}. 
In the following, we briefly review this formalism and extend it to describe {\em AC} Josephson transport. 
The system under consideration can be divided into three subsystems, the dot, the fermionic states of the leads, and the Cooper pair condensates in the superconductors. 
Since we are not interested in the fermionic dynamics of the leads, we can trace out their degrees of freedom and arrive at a reduced density matrix for the remaining part, i.e., the dot's degrees of freedom and the Cooper pair condensates. 
Its elements are given by $P^{\xi_1}_{\xi_2}\equiv \bra{\xi_1}\rho_\text{red}\ket{\xi_2}$, where $\ket{\xi}\equiv\ket{\chi,\{n_{S1},n_{S2}\}}$ includes the dot state $\chi=0,\uparrow,\downarrow,d$ as well as the number of Cooper pairs in the two superconductors, $n_{S1}$ and $n_{S2}$, measured relative to an arbitrary but fixed reference. 
The diagonal elements $P_\xi\equiv P_\xi^\xi$ give the probability to be in state $\xi$. 
The off-diagonal elements $P^{\xi_1}_{\xi_2}$ with $\xi_1\neq \xi_2$ describe coherent superpositions.
The states $\xi_1$ and $\xi_2$ in $P^{\xi_1}_{\xi_2}$ provide more information than is needed to study the electric transport. 
In fact, only the {\em differences} of the Cooper pair numbers of the states $\ket{\xi_1}$ and $\ket{\xi_2}$ are important.
Moreover, particle number conservation sets the constraint that the total number of electrons in $\ket{\xi_1}$ has to be the same as in $\ket{\xi_2}$.
For convenience, we define
\begin{equation}
	P^{\chi_1}_{\chi_2}(\{n_{S1},n_{S2}\})\equiv \sum\limits_{m_{S1},m_{S2}} P^{(\chi_1,\{m_{S1}+n_{S1},m_{S2}+n_{S2}\})}_{(\chi_2,\{m_{S1},m_{S2}\})}\, .
\end{equation}
From the definition it follows that the symmetry relation $P^{\chi_1}_{\chi_2}(\{n_{S1},n_{S2}\}) = \left[ P^{\chi_2}_{\chi_1}(\{-n_{S1},-n_{S2}\}) \right]^*$ holds.
As a consequence of particle conservation, $n_{S2}$ is a unique function of $\chi_1$, $\chi_2$, and $n_{S1}$. 
It is, therefore, enough to keep track of the Cooper-pair number $n$ of one lead only.
We choose here lead $S1$, i.e., $n=n_{S1}$, and introduce the definitions $P_\chi(n)\equiv P_\chi^\chi(\{n,-n\})$ and
$P^d_0(n)\equiv P^d_0(\{n,-n-1\})$.
To inherit the symmetry relation $P^0_d(n) = \left[ P^d_0(-n) \right]^*$, we consistently define $P^0_d(n)\equiv P^0_d(\{n,-n+1\})$. 

Finally, we collect all the nonvanishing elements of the reduced density matrix in the vector ${\bs \pi}(n)\equiv (P_0(n),P_\uparrow(n),P_\downarrow(n),P_d(n),P^{d}_{0}(n),P^{0}_{d}(n))^T$. Its dynamics is governed by the generalized master equation
\begin{align}
\label{genmaeq}
	&\frac{\text{d}}{\text{d}t}{\bs \pi}(n)(t)+\text{i}{\bf E}_n{\bs \pi}(n)(t)
	\nonumber \\
	=&\sum\limits_{n'}\int \limits_{-\infty}^t\mathrm{d}t'{\bf W}(n,n')(t,t'){\bs \pi}(n')(t')\, ,
\end{align}
where the matrix elements of the kernel $\left.W\right.^{\chi_1\chi_1'}_{\chi_2\chi_2'}(n,n')(t,t')$ are the transition rates from an initial state at time $t'$ described by $P^{\chi_1'}_{\chi_2'}(n')(t')$ to a final state at time $t$ described by $P^{\chi_1}_{\chi_2}(n)(t)$. 
For the kernel we have introduced a notation analogous to the one adopted for the reduced density matrix: $W^{\chi_1\chi_1'}_{\chi_2\chi_2'}(n_{S1},n'_{S1})(t,t')\equiv W^{\chi_1\chi_1'}_{\chi_2\chi_2'}(\{n_{S1},n_{S2}\},\{n'_{S1},n_{S2}'\})(t,t')$, where the excess number of Cooper pairs in the superconductor $S2$ is fixed by particle conservation.  
We get $n_{S1} +n_{S2}$ to be equal to $0$ for $\chi_1=\chi_2$, equal to $-1$ for $\chi_1=d$, $\chi_2=0$, and equal to $+1$ for $\chi_1=0$, $\chi_2=d$. In a similar way, $n_{S2}'$ is determined in terms of $n_{S1}'$, $\chi_1'$, and $\chi_2'$.  
The only nonvanishing matrix elements of the matrix ${\bf E}_n$ are $\left.E_n\right.^{\chi\chi}_{\chi\chi}= 2neV$, $\left.E_n\right.^{dd}_{00}= \delta_{S2}+2neV=\delta_{S1}+2(n+1)eV$, and $\left. E_{n}\right.^{00}_{dd}=\left. -E_{-n}\right.^{dd}_{00}$, where $V$ is the voltage drop between $S2$ and $S1$ and it reads $-eV=\mu_{S2}-\mu_{S1}$ with $e>0$.
 
The tunneling current between the dot and lead~$r$ is given by
\begin{equation}
\label{current}
	I_r(t)=e\sum\limits_{n'}\int \limits_{-\infty}^t\mathrm{d}t'{\bf e}^T{\bf W}^r(0,n')(t,t'){\bs \pi}(n')(t')\, ,
\end{equation}
where ${\bf e}^T=(1,1,1,1,0,0)$ and $n=0$ in ${\bf W}^r(0,n')(t,t')$ ensure that the final state on the right hand side is diagonal both in the dot state and the Cooper pair numbers.
The current rates ${\bf W}^{r}(n,n')(t,t')$ are similar to the general rates ${\bf W}(n,n')(t,t')$ but take into account the electrons transferred from lead $r$ to the dot.
In addition to tunneling currents, there are, in general, displacement currents due to the formation of image charges when the dot occupation varies in time.\cite{bruder94}
The displacement currents do not play any role for the {\em DC} part.
But even for the {\em AC} part they drop out for the symmetrized current $I_S(t) \equiv [I_{S1}(t) - I_{S2}(t)]/2$ when choosing the capacitances of the tunnel contacts between the dot and the two superconductors symmetrically. 
Therefore, we ignore the displacement currents in the following.

The frequency of the {\em AC} Josephson signal is given by the energy difference of a Cooper pair being in  superconductor $S1$ or $S2$, i.e., by $2(\mu_{S1}-\mu_{S2})=2 eV$.
Therefore, we perform a Fourier expansion by making use of $A(t)=\sum_{n=-\infty}^{\infty}A^n e^{2\m{i}neV t}$ and $A^n=(1/\mathcal{T})\int_0^\mathcal{T}\m{d}tA(t)e^{-2\m{i}neVt}$ with $\mathcal{T}=2\pi/(2eV)$.
Within the diagrammatic approach the factor $\exp(-2\m{i}neVt)$ appearing in the $n$-th Fourier component of the current simply adds a term $2neV$ to the energy difference of the states on the upper and lower Keldysh contour.
This term can easily be incorporated into the energy difference arising from different Cooper pair numbers by shifting 
${\bf W}^r(0,n')(t,t')$ and ${\bs \pi}(n')(t')$ in Eq.~(\ref{current}) to ${\bf W}^r(n,n'+n)(t,t')$ and ${\bs \pi}(n'+n)(t')$, respectively, i.e., only the $0$-th Fourier components of ${\bf W}^{r}$ and ${\bs \pi}$ are needed.
Performing the remaining time integral, we get for the $n$-th Fourier component of the current
\begin{equation}
\label{fouriercurrent}
	I_r^n=e\sum\limits_{n'}{\bf e}^T{\bf W}^{r}(n,n'){\bs \pi}(n')\, ,
\end{equation}
with ${\bf W}=\int_{-\infty}^t\m{d}t'{\bf W}(t,t')e^{\m{i}0^+t'}$ being the zero frequency Laplace transformed rate, that does not depend on the final time $t$. 
The $0$-th Fourier components ${\bs \pi}(n')$ are readily obtained from the $0$-th Fourier component of Eq.~(\ref{genmaeq}),
\begin{eqnarray}
\label{masterequcooper}
\mathrm{i}{\bf E}_n{\bs \pi}(n)=\sum\limits_{n'}{\bf W}(n,n'){\bs \pi}(n') 
\end{eqnarray}
together with the normalization condition ${\bf e}^T{\bs \pi}(n)=\delta_{n,0}$.
In summary, the $n$-th Fourier component of the current can be evaluated within the diagrammatic technique in exactly the same way as the {\em DC} current (see Ref.~\onlinecite{governale08}) but allowing for off-diagonal final Cooper pair states ($n\neq 0$) in ${\bf W}^{r}(n,n')$ on the right hand side of Eq.~(\ref{fouriercurrent}).
The diagrammatic rules to calculate the kernels ${\bf W}$ and ${\bf W}^r$ are given in Appendix~\ref{appdiag}.
 
\section{results}
In the following, we perform a systematic perturbation expansion of ${\bs \pi}(n)$, ${\bs W}(n,n')$, ${\bs W}^r(n,n')$ and $I_r^n$ in the tunnel-coupling strengths, $\Gamma\equiv \text{max}\{\Gamma_{S1},\Gamma_{S2},\Gamma_N\}$. 
Since we assume the tunnel couplings to be weak, we restrict ourselves to lowest (first) order for the kernels ${\bs W}(n,n')$ and ${\bs W}^r(n,n')$.
In addition, we concentrate on the limit of an infinite superconducting gap in the leads, $\Delta\rightarrow \infty$, i.e.,  quasi-particle tunneling between dot and the superconductors is suppressed.
As a consequence, the current into the superconductors is exclusively sustained by Cooper pairs. 
The normal lead affects the occupation of the quantum dot, which, in turn, affects Cooper-pair transport.
Even a weakly tunnel-coupled normal conductor influences {\em AC} Josephson transport between the two superconductors. 

For $\Delta \rightarrow \infty$ and to first order in $\Gamma$, all matrix elements of ${\bf W}(n,n')$ and ${\bf W}^r(n,n')$ that require either higher-order tunneling or a finite superconducting gap $\Delta$ in the leads vanish. 
The only non-vanishing ones entering Eq.~(\ref{fouriercurrent}) are readily evaluated, see Appendix~\ref{app_rates}.
This results in
\begin{equation}
\label{current-gen-form}
 	I_{S1}^n = \m{i}e\Gamma_{S1} \left[P^0_d (n+1) -P^d_0 (n-1) \right] 
	\, ,
\end{equation}
i.e., the $n$-th component to the current into superconductor $S1$ is fully determined by the density matrix elements $P^d_0 (n-1)$ and $P^0_d (n+1)=\left[ P^d_0 (-n-1) \right]^*$.
The latter describe superconducting correlations induced on the quantum dot due to the proximity effect. 

The Cooper pair degree of freedom $n$ in ${\bs \pi}(n)$ introduces an apparently infinitely large number of density matrix elements that are all coupled to each other via Eq.~(\ref{masterequcooper}).
However, in the limit of a large bias voltage as compared to the tunnel-coupling strength, $|eV|\gg \Gamma$, only very few of them need to be taken into account.
This is a consequence of ${\bf E}_n$ appearing on the left hand side of Eq.~(\ref{masterequcooper}).
Most of its matrix elements are of order $eV$, while ${\bf W}(n,n')$ on the right hand side scales with $\Gamma$.
This mismatch defines a hierarchy in powers of $\Gamma/(eV)$ for the density matrix elements.
The lowest order contains all matrix elements of ${\bs \pi}(n)$ for which the corresponding ${\bf E}_n$ is zero or of the order of $\Gamma$.
This includes all diagonal matrix elements $P_\chi^\chi(0)$ (and excludes all elements $P_\chi^\chi(n)$ with $n\neq 0$).
The next order contains all matrix elements of ${\bs \pi}(n)$, that can be connected to lowest order ones by the kernel ${\bf W}(n,n')$.
The only off-diagonal matrix elements that can be reached from the diagonal ones for $\Delta \rightarrow \infty$ and to first order in $\Gamma$ are $P^d_0 (-1) = \left[ P^0_d (1) \right]^*$ and $P^d_0 (0) = \left[P^0_d (0)\right]^*$. 
If the gate voltage is tuned such that the quantum dot is in resonance either with superconductor $S1$ or $S2$, namely $|\delta_{S1}| \lesssim \Gamma$ or $|\delta_{S2}| \lesssim \Gamma$, then $P^d_0 (-1)$ or $P^d_0 (0)$ already belong to the lowest order in the hierarchy, indicating strong proximity effect with superconductor $S1$ or $S2$, respectively.
But in any case, all off-diagonal matrix elements except $P^d_0 (-1) = \left[ P^0_d (1) \right]^*$ and $P^d_0 (0) = \left[P^0_d (0)\right]^*$ can be dropped for describing the current into the superconductors. 

Starting with the kinetic equations with off-diagonal final states,
\begin{align}
\label{eqdiag1}
	\m{i}(\delta_{S2}+2neV) P^d_0(n) =& W_{0 0}^{d 0} (n,0)P_0 (0) + W_{0 d}^{d d} (n,0)P_d (0)
	\nonumber \\
	&+W^{d d}_{0 0} (n,n) P_0^d (n)
	\, ,
\end{align}  
for $n=-1,0$,
and using the rates listed in Appendix~\ref{app_rates}, we find that the required off-diagonal density matrix elements are related to the diagonal ones via
\begin{subequations}
\begin{eqnarray}
	P^d_0 (-1)&=&\frac{\Gamma_{S1}}{2 A_{S1}} \left[ P_0(0)-P_d(0) \right]
\label{Poff(-1)}
\\
	P^d_0 (0)&=&\frac{\Gamma_{S2}}{2 A_{S2}} \left[ P_0(0)-P_d(0) \right]
\label{Poff(0)}
\, ,
\end{eqnarray}
\end{subequations}
where we defined the complex resolvents $A_{S1}=\delta_{S1}+\m{i} W_{00}^{dd}(-1,-1)$ and $A_{S2}=\delta_{S2}+\m{i} W_{00}^{dd}(0,0)$.
The expressions for $W_{00}^{dd}(-1,-1)$ and $W_{00}^{dd}(0,0)$ are given in Appendix~\ref{app_rates}.
Their imaginary parts can be interpreted as the renormalization of the detuning $\delta_{S1}$ and $\delta_{S2}$, respectively, due to the tunnel coupling to the normal lead.
Their real parts provide a width to the resonances. 
For a systematic perturbation expansion, we may replace the full expressions for the resolvents $A_r$ by their leading-order term only.
To do so, we need to distinguish the two cases of the quantum dot to be on or off resonance with superconductor $r$.
On resonance, $|\delta_r| \lesssim \Gamma$, we find that $A_r$ starts to first order, i.e., we can omit the $\delta_r$ appearing in the argument of the Fermi and the digamma functions to arrive at ${\rm Re} \, A_r = \delta_{r} + \sigma$ with $\sigma\equiv \frac{\Gamma_N}{\pi}\m{Re}\left[\Psi\left(\frac{1}{2}+\m{i}\frac{\e+U-\mu_N}{2\pi k_BT}\right)-\Psi\left(\frac{1}{2}+\m{i}\frac{\e-\mu_N}{2\pi k_BT}\right)\right]$ and ${\rm Im} \, A_r = -\Gamma_N\left[1+f(\e)-f(\e+U) \right]$.
Off resonance, $|\delta_r| \gg \Gamma$, we can replace the resolvent by its zeroth-order term, $A_r = \delta_{r}$.

In order to determine the non-vanishing elements of the reduced density matrix we also need the kinetic equations with diagonal final states,
\begin{align}
\label{dotoccudetail}
	0=& \sum_{\chi'}W_{\chi \chi'}^{\chi \chi'} (0,0)P_{\chi'} (0) 
	\nonumber \\
	&+2  \sum_{n'=-1,0} {\rm Re} \left[ W^{\chi d}_{\chi 0} (0,n') P_0^d (n') \right] 
	\, .
\end{align}
The rates with diagonal initial and finite states  are related to single-electron tunneling between dot and normal conductor. In contrast, the rates connecting superpositions between a doubly-occupied and an empty dot to a diagonal state require tunneling of one Cooper pair from or to the condensate of a superconducting lead.
As a result, we find
\begin{align}
	P_0(0)-P_d(0) = \frac{1-f(\varepsilon)-f(\varepsilon+U)}{1+f(\varepsilon)-f(\varepsilon+U)+  {\rm Im} \sum\limits_{r=S1,S2}  \frac{\Gamma_r^2}{\Gamma_N A_r}}
\end{align}
Plugging this into Eqs.~(\ref{Poff(-1)}) and (\ref{Poff(0)}) and employing Eq.~(\ref{current-gen-form}) yields the current into superconductor $S1$ for all values of $\e$.
This is what we use to calculate all the curves in the figures.
The resulting formulae can be simplified further to obtain compact analytical results after specifying whether the quantum dot is in resonance with one of the superconductors or not.

First, we observe from Eq.~(\ref{current-gen-form}) that only the zeroth and first Fourier component of the current flowing into superconductor $S1$ are nonvanishing.
For the {\em DC} current, $I_{S1}^{DC} = I_{S1}^{0} = 2e \Gamma_{S1} {\rm Im }\, P_0^d(-1)$, it is important whether the dot is in resonance with the same superconductor the current is measured in.
On resonance, $|\delta_{S1}| \lesssim \Gamma$, we find
\begin{equation}
	\label{S1statio}
	I^{DC}_{S1}=\frac{e\Gamma_N\Gamma_{S1}^2 \left[ 1-f(\e)-f(\e+U)\right]}{(\delta_{S1}+\sigma)^2+\Gamma_{S1}^2 + \Gamma_N^2 \left[ 1+f(\e)-f(\e+U)\right]^2}\, ,
\end{equation}
which starts in first order in $\Gamma$.
Off resonance, the {\em DC} current starts only in third order in $\Gamma$, i.e., vanishes to the order considered here. 
In summary, a {\em DC} current flows only between the normal lead and the superconductor which is in resonance with the quantum dot.
We explicitly checked that $I_{S1}^{DC}=-I^{DC}_{N}$, which guarantees current conservation.

We now turn to the {\em AC} current $I_{S1}^{AC}(t)$.
It can be decomposed into a cos- and a sin-term, $I_{S1}^{AC}(t) = I_{S1}^{AC, \rm cos} \cos ( 2eV t) +  I_{S1}^{AC, \rm sin} \sin ( 2eV t)$ with $I_{S1}^{AC, \rm cos} = I_{S1}^{1}+I_{S1}^{-1}=2e \Gamma_{S1} {\rm Im }\, P_0^d(0)$ and $I_{S1}^{AC, \rm sin} = \m{i}\left( I_{S1}^{1}-I_{S1}^{-1}\right)=2e \Gamma_{S1} {\rm Re }\, P_0^d(0)$.
The current $I_{S2}^{AC}(t)$ in $S2$ can be obtained from $I_{S1}^{AC}(t)$ by replacing $\Gamma_{S1} \leftrightarrow \Gamma_{S2}$ and $\mu_{S1}\leftrightarrow \mu_{S2}$.

We immediately see that the behavior of the {\em AC} tunneling current into one superconductor depends on whether the quantum dot is in resonance with the {\em other} superconducting lead, which supports the interpretation that, to lowest order in $\Gamma$, the {\em AC} tunneling current into superconductor $S1$ is sustained by oscillations of Cooper pairs between lead $S1$ and the quantum dot that is proximized by the tunnel coupling to lead $S2$.

Similarly as the {\em DC} current into lead $S1$, the cos-part of the {\em AC} current starts in third order in $\Gamma$ as long as the quantum dot is off resonance with lead $S2$.
On resonance, $|\delta_{S2}| \lesssim \Gamma$, we obtain
\begin{equation}
	I^{AC,\rm cos}_{S1}=\frac{e\Gamma_N\Gamma_{S1}\Gamma_{S2} \left[ 1-f(\e)-f(\e+U)\right]}{(\delta_{S2}+\sigma)^2+\Gamma_{S2}^2 + \Gamma_N^2 \left[ 1+f(\e)-f(\e+U)\right]^2}\, .
\end{equation}
The sine-part, on the other hand, is on resonance, $|\delta_{S2}| \lesssim \Gamma$, given by 
\begin{eqnarray}
	I^{AC,\rm sin}_{S1}&=&\frac{e\Gamma_{S1}\Gamma_{S2} \left( \delta_{S2} + \sigma \right)}{(\delta_{S2}+\sigma)^2+\Gamma_{S2}^2 + \Gamma_N^2 \left[ 1+f(\e)-f(\e+U)\right]^2}
	\nonumber \\
	&&\times \frac{1-f(\e)-f(\e+U)}{1+f(\e)-f(\e+U)}
	\, ,
\end{eqnarray}
while off resonance, $|\delta_{S2}| \gg \Gamma$, we find
\begin{equation}
	I^{AC,\rm sin}_{S1}=\frac{e\Gamma_{S1}\Gamma_{S2}}{\delta_{S2}}\cdot 
	\frac{1-f(\e)-f(\e+U)}{1+f(\e)-f(\e+U)} \, .
\end{equation}
It is remarkable that the {\em amplitude} and {\em phase} of the {\em AC} tunneling current into superconductor $S1$ does not depend on the chemical potential $\mu_{S1}$ but only on the detuning $\delta_{S2}$ between quantum dot and the otherÊ superconductor $S2$.
The chemical potential $\mu_{S1}$ only enters the oscillation {\em frequency}, given by the bias voltage $2eV=2(\mu_{S1}-\mu_{S2})$. 
This can be interpreted in the following way: the tunnel coupling to the superconducting leads induces superconducting correlations on the quantum dot. 
The total proximity effect is given by the sum of the contributions stemming from the two superconductors. 
The {\em AC} Josephson current between quantum dot and $S1$, however, only probes the proximity effect induced by superconductor $S2$, held at a different chemical potential.

Another interesting feature of the expression for the {\em AC} current is the behavior for $V\rightarrow0$. 
The amplitude of the sine first harmonic exactly reproduces the {\em DC} Josephson transport between the two superconductors that was discussed in Ref.~\onlinecite{pala07}. 
This crossover has been discussed before in quantum point contacts\cite{averin95,jonckheere09} as well as in the three-terminal setup under consideration in this paper in the noninteracting limit.\cite{jonckheere09} 
We also notice that the cosine term, $I^{AC,\rm cos}_{S1}$, vanishes for $\Gamma_N\rightarrow 0$ while the sine term remains finite.

In the following, we discuss the amplitude $|I_S^{AC}|$ and the phase $\phi$ of only the symmetrized current $I_S^{AC}(t)= [ I_{S1}^{AC}(t) - I_{S2}^{AC}(t)]/2 = |I_S^{AC}| \sin(2eVt + \phi)$, for which the displacement currents drop out. 
Without loss of generality, we choose the reference energy such that $\mu_N=0$.
Furthermore, we concentrate on the limit of symmetric tunnel coupling to the two superconductors, $\Gamma_{S1}=\Gamma_{S2} \equiv \Gamma_S$.
We distinguish the two cases of $\Gamma_N \approx \Gamma_S$ and $\Gamma_N \ll \Gamma_S$.

\begin{figure}
\includegraphics[width=.95\columnwidth]{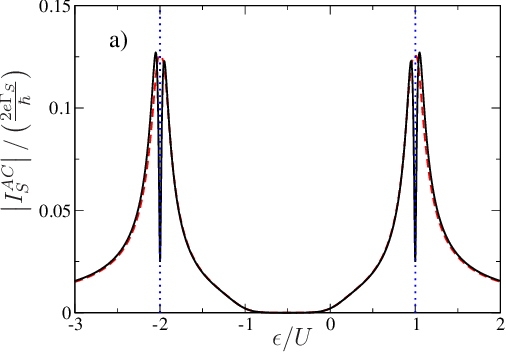}\\[2ex]
\includegraphics[width=.95\columnwidth]{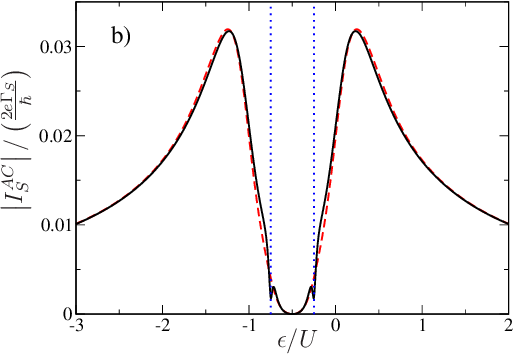}
\caption{
\label{amplitude_symm}
Amplitude of the {\em AC} Josephson current in units of $2e\Gamma_S/\hbar$ as a function of dot level position $\e/U$.
The temperature is $k_BT = U/10$, the bias voltage to the normal lead is $\mu_N=0$, and the bias voltages $\mu_{S1}$ and $\mu_{S2}$ applied to the superconductors are, respectively, given by a) $-3U/2$ and $3U/2$, b) $-U/4$ and $U/4$.
The solid curves are for $\Gamma_N= \Gamma_S/10=U/100$, the dashed lines for $\Gamma_N=\Gamma_S=U/10$.
The resonance conditions are indicated by vertical lines.
}
\end{figure}

\begin{figure}
\includegraphics[width=.95\columnwidth]{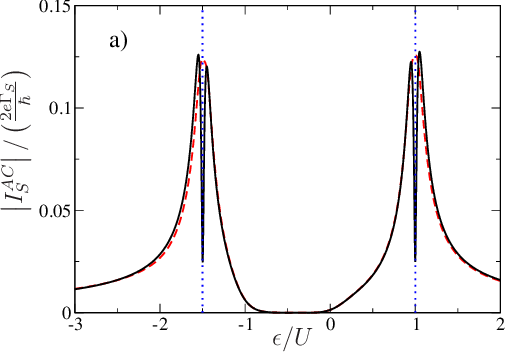}\\[2ex]
\includegraphics[width=.95\columnwidth]{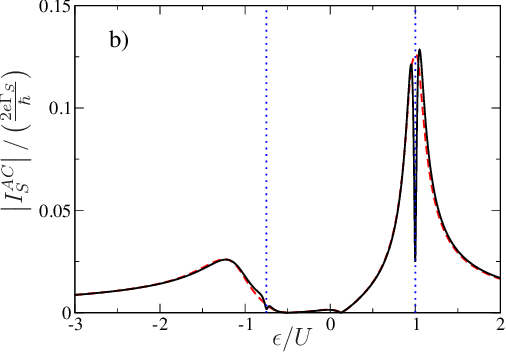}\\[2ex]
\includegraphics[width=.95\columnwidth]{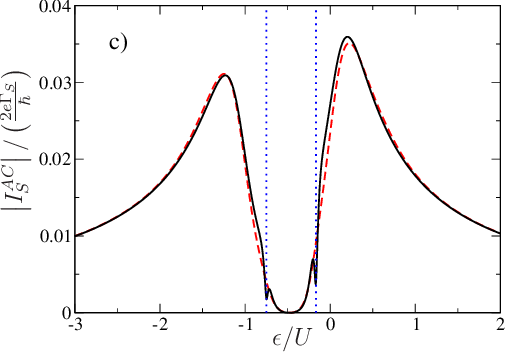}\\[2ex]
\caption{
\label{amplitude_asymm}
Amplitude of the {\em AC} Josephson current in units of $2e\Gamma_S/\hbar$ as a function of dot level position $\e/U$.
The temperature is $k_BT = U/10$, the bias voltage to the normal lead is $\mu_N=0$, and the bias voltages $\mu_{S1}$ and $\mu_{S2}$ applied to the superconductors are, respectively, given by  a) $-U$ and $3U/2$, b) $-U/4$ and $3U/2$, and c) $-U/4$ and $U/3$.
The solid curves are for $\Gamma_N= \Gamma_S/10=U/100$, the dashed lines for $\Gamma_N=\Gamma_S=U/10$.
The resonance conditions are indicated by vertical lines. }
\end{figure}

In Figs.~\ref{amplitude_symm} and Figs.~\ref{amplitude_asymm}, we plot the amplitudes of the {\em AC} Josephson currents as a function of the dot level position $\e$ for five different values of the applied bias voltages $\mu_{S1}$ and $\mu_{S2}$. Thereby, the voltage is applied either symmetrically or asymmetrically with respect to $\mu_N$.
Coulomb blockade suppresses the {\em AC} Josephson effect for $-U < \e < 0$, since in this region the quantum dot is predominantly singly occupied. 
This suppression is also present for {\em DC} Josephson transport through the quantum dot, as has been observed experimentally\cite{vandam06,jorgensen07} and discussed theoretically.\cite{governale08}
Outside this region, an occupation of the quantum dot with an even number of electrons is possible and, hence, Josephson transport is possible.

The dot level energies $\e$ around which the resonance conditions $|\delta_{S1}| \lesssim \Gamma$ and $|\delta_{S2}| \lesssim \Gamma$ to be in resonance with superconductor $S1$ and $S2$ are fulfilled are given by $\e \approx \mu_{S1} -U/2$ and $\e \approx \mu_{S2} -U/2$, respectively. Around these points, which are indicated in the figure by dotted lines, the AC tunneling current between quantum dot and the {\em other} superconductor, $S2$ and $S1$, respectively, is enhanced. 
Depending on the biasing, each of the two resonances lie either inside or outside the Coulomb-blockade gap. In the first case, AC Josephson transport is suppressed at the resonances and the amplitude of the {\em AC} Josephson current shows a local maximum close to the edge of the Coulomb-blockade gap. In the latter case, a maximimum is clearly visible whenever the dot is resonance with one of the superconductors.

In Figs.~\ref{amplitude_symm} a) and \ref{amplitude_asymm} a), both resonances are outside, in Fig.~\ref{amplitude_asymm} b) one is outside and one inside, and in Figs. ~\ref{amplitude_symm} b) and ~\ref{amplitude_asymm} c) both are inside the Coulomb-blockade region.

\begin{figure}
\includegraphics[width=.95\columnwidth]{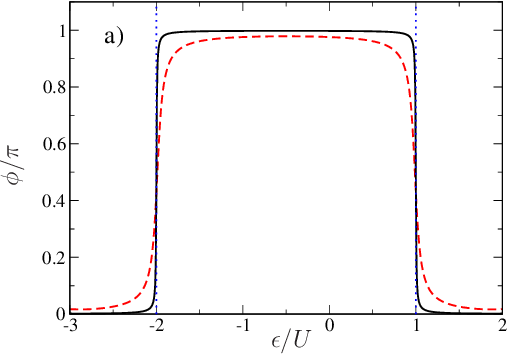}\\[2ex]
\includegraphics[width=.95\columnwidth]{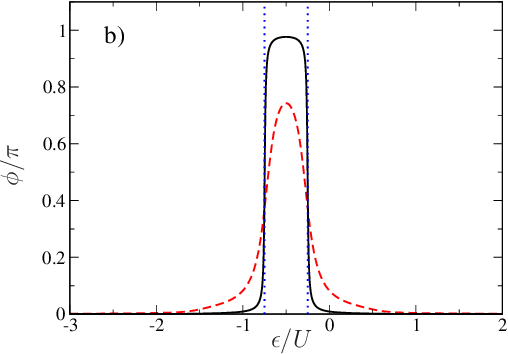}
\caption{
\label{phase}
Phase of the {\em AC} Josephson current in units of $\pi$ as a function of dot level position $\e/U$.
The parameters are the same as in Fig.~\ref{amplitude_symm} a) and b).
}
\end{figure}

The phase of the {\em AC} Josephson current for symmetrically applied bias voltages is shown in Fig.~\ref{phase}.
We find that there is $0$-$\pi$ transition when crossing through the resonances.
This means that away from the resonances, only the sin-term contributes to the {\em AC} Josephson current, while the cos-term vanishes.
At the resonance, however, the sin-term changes sign, i.e., goes through zero, while the cos-term remains finite.
For $\Gamma_N = \Gamma_S$, dashed lines in Figs.~\ref{amplitude_symm} and \ref{amplitude_asymm} , this $0$-$\pi$ transition is not visible in the amplitude but only in the phase (the phase corresponding to the asymmetric case, Fig. ~\ref{amplitude_asymm}, is not shown).  
In contrast, for $\Gamma_N \ll \Gamma_S$, solid lines in Figs.~\ref{amplitude_symm} and \ref{amplitude_asymm}, the $0$-$\pi$ transition is indicated by a sharp dip in the amplitude as well.
This is a consequence of the fact, that the cos-term starts linearly in $\Gamma_N$ while the sin-term remains finite even for $\Gamma_N \rightarrow 0$.
 
\section{conclusions}
We have analyzed {\em AC} Josephson transport through a single-level quantum dot tunnel coupled to two superconductors and a normal conductor. 
The amplitude and the phase of the {\em AC} Josephson current depend both on the gate and bias voltages.
As a function of gate voltage, there are two resonances:
When the quantum dot is in resonance with one superconductor, the {\em AC} Josephson current between dot and the {\em other} superconductor is enhanced.
There is a $0$-$\pi$ transition at each of the resonances.
For small tunnel coupling to the normal lead, this $0$-$\pi$ transition is accompanied with a sharp dip in the amplitude of the {\em AC} Josephson current.
Inside the Coulomb-blockade region, i.e., the region of gate voltages for which the quantum dot is predominantly singly occupied, the {\em AC} Josephson current is exponentially suppressed. 
The frequency of the {\em AC} Josephson oscillations is given by the voltage difference between the two superconductors.

\textit{Acknowledgments}. We acknowledge financial support from DFG via KO 1987/5. 

\appendix

\section{Diagrammatic Rules} 
\label{appdiag}
The rules for evaluating the generalized rates $\left.W\right.^{\chi_1\chi_1'}_{\chi_2\chi_2'}(\{n_1,n_2\},\{n_1',n_2'\})$ are as follows:\\

(1) Draw all topologically different diagrams with fixed ordering of 
the vertices in the real axis. 
The vertices are connected in pairs by tunneling lines carrying energy 
$\omega_i$.
The tunneling lines can be normal or anomalous. 
For each anomalous line choose the direction (forward or backward with respect 
to the Keldysh contour) arbitrarily.

(2) For each vertical cut between two vertices assign a factor 
$1/(\Delta E +i \eta)$ with $\eta=0^+$, where $\Delta E$
is the difference between the left-going and the right-going energies,
including the energy of the dot states $E_\chi$, the tunneling lines 
$\omega_i$, and the energy difference in Cooper-pair condensates 
$E_{\text{CP}}$.
The latter is increased (decreased) at each vertex of an outgoing (incoming)
anomalous line at which the arrow is opposite to the arbitrarily chosen line 
direction.

(3) For each tunneling line assign a factor
$\frac{1}{2 \pi} \Gamma_r D_r(\omega_i) f^\pm_r(\omega_i)$,
where $f^+_r(\omega_i) = f_r(\omega_i) = [1+\exp(\omega_i-\mu_r)/ (k_{{\rm B}} T)]^{-1}$ and $f^-_r(\omega_i) = 
1 - f_r(\omega_i)$, and
$D_r(\omega_i)=\frac{|\omega_i-\mu_r|}{\sqrt{(\omega_i-\mu_r)^2 -
|\Delta_r|^2}} \theta(|\omega-\mu_r |-|\Delta_r|)$.
The upper (lower) sign applies for lines going backward (forward) with respect 
to the Keldysh contour.
For anomalous lines multiply an additional factor \footnote{In Ref. \onlinecite{governale08} this factor contained a typo.} 
$\pm \text{sign}(\omega_i-\mu_r) \frac{|\Delta_r|}{|\omega_i-\mu_r|}$. Moreover, assign a factor $e^{-\m{i} \Phi_r}$ for an outgoing and
$e^{\m{i} \Phi_r}$ for an incoming anomalous line.
[For normal leads, only normal lines with $D_r(\omega_\m{i}) \equiv 1$ appear.]

(4) Assign an overall prefactor $-\m{i}$.\\
Furthermore, assign a factor $-1$ for each\\
a) vertex on the lower propagator;\\
b) crossing of tunneling lines;\\
c) vertex that connects the doubly-occupied dot state, 
$| d \rangle = d^\dagger_{\uparrow} d^\dagger_{\downarrow} | 0 \rangle$, 
to spin up, $| \uparrow \rangle$;\\
d) outgoing (incoming) anomalous tunneling line in which the earlier (later)
tunnel vertex with respect to the Keldysh contour involves a spin-up
dot electron.\\ \protect
[The factors in c) and d) arise due to Fermi statistics from the order of 
the dot and lead operators, respectively.]

(5) For each diagram, integrate over all energies $\omega_i$.
Sum over all diagrams.\\

The generalized current rates \footnote{Since in Ref. \onlinecite{governale08} only stationary currents have been discussed only rules for $W^{\chi \chi_1' r}_{\chi \chi_2'}(\{0,0\},\{n_1',n_2'\})$ have been presented. These are also applicable to arbitrary current rates $W^{\chi \chi_1' ,r}_{\chi \chi_2'}(\{n_1,n_2\},\{n_1',n_2'\})$.} 
$W^{\chi \chi_1,r}_{\chi \chi_2'}(\{n_1,n_2\},\{n_1',n_2'\})$
are evaluated in the following way:\\

(6) Multiply the value of the corresponding generalized rate 
$W^{\chi \chi_1'}_{\chi \chi_2'} (\{n_1,n_2\},\{n_1',n_2'\})$ with a factor
given by adding up the following numbers for each tunneling 
line that is associated with lead $r$:\\
a) for normal lines: $1$ if the line is going from the lower to the 
upper, $-1$ if it is going from the upper to the lower propagator, and $0$
otherwise;\\
b) for anomalous lines: $1$ for incoming lines within the upper propagator and
outgoing lines within the lower propagator, $-1$ for outgoing lines within 
the upper propagator and incoming lines within the lower propagator, and $0$ otherwise.

\section{Rates and Current Rates}
\label{app_rates}

In this appendix, we list all the rates and current rates entering the calculation. 
Thereby, we omit all rates that can be obtained from the listed ones via the symmetry relations
$W^{\chi_1 \chi_1'}_{\chi_2 \chi_2'} ( n, n') = \left[ W^{\chi_2 \chi_2'}_{\chi_1 \chi_1'} ( -n, -n') \right]^*$ and
$W^{\chi_1 \chi_1',r}_{\chi_2 \chi_2'} ( n, n') = \left[ W^{\chi_2 \chi_2',r}_{\chi_1 \chi_1'} ( -n, -n') \right]^*$.

\subsection{Superconductor}
 
To lowest order in $\Gamma$ and for $\Delta \rightarrow \infty$, many rates and currents rates involving a superconducting tunneling line vanish.
The non-vanishing ones turn out to be independent of the Cooper pair numbers of the condensates. 
We find
\begin{align}
&W^{0 d}_{0 0}(n,n-1)=W^{d 0}_{0 0}(n,n+1)=\m{i}\Gamma_{S1}/2
\nonumber \\
&W^{d d}_{d 0}(n,n-1)=W^{d d}_{0 d}(n,n+1)=-\m{i}\Gamma_{S1}/2
\nonumber \\
&W^{0 d}_{0 0}(n,n)=W^{d 0}_{0 0}(n,n)=\m{i}\Gamma_{S2}/2
\nonumber \\
&W^{d d}_{d 0}(n,n)=W^{d d}_{0 d}(n,n)=-\m{i}\Gamma_{S2}/2
\nonumber 
\, ,
\end{align}
for the rates and
\begin{align}
&W^{0 d, {S1}}_{0 0}(n,n-1)=W^{d d, {S1}}_{d 0}(n,n-1)=-\m{i}\Gamma_{S1}/2
\nonumber \\
&W^{0 d, {S2}}_{0 0}(n,n)=W^{d d, {S2}}_{d 0}(n,n)=-\m{i}\Gamma_{S2}/2
\nonumber  
\, ,
\end{align}
for the current rates.

\subsection{Normal Conductor}

Changing the state of the dot due to tunneling from and to the normal conductor is described by the rates
$W_{\chi \chi'}(0,0) \equiv W^{\chi \chi'}_{\chi \chi'}(0,0)$ with
\begin{align}
&W_{\sigma 0}(0,0)=\Gamma_N f(\e)\nonumber \\
&W_{0\sigma}(0,0)=\Gamma_N [1-f(\e)]\nonumber \\
&W_{d\sigma}(0,0)=\Gamma_N f(\e+U)\nonumber \\
&W_{\sigma d}(0,0)=\Gamma_N [1-f(\e+U)]\nonumber \, ,
\end{align}
and $W_{0 0}(0,0)=-2W_{\sigma 0}(0,0)$, $W_{\sigma \sigma}(0,0)= -W_{0 \sigma}(0,0)-W_{d \sigma}(0,0)$, as well as $W_{d d}(0,0)=-2W_{\sigma d}(0,0)$.
The nonvanishing current rates are given by $W_{\chi \chi'}^N(0,0) \equiv W^{\chi \chi',N}_{\chi \chi'}(0,0)$ with 
\begin{align}
&W_{\sigma 0}^N(0,0)=\Gamma_N f(\e)\nonumber \\
&W_{0\sigma}^N(0,0)=- \Gamma_N [1-f(\e)]\nonumber \\
&W_{d\sigma}^N(0,0)=\Gamma_N f(\e+U)\nonumber \\
&W_{\sigma d}^N(0,0)=-\Gamma_N [1-f(\e+U)]\nonumber \, .
\end{align}

Finally, we need the kernels $W^{d d}_{0 0}(n,n)$ for $n=-1,0$. 
For $n=-1$, we find
\begin{align}
&W^{d d}_{0 0}(-1,-1)=-\Gamma_N\left[1+f(\e-\delta_{S1})-f(\e+U-\delta_{S1}) \right] \nonumber \\
&+\frac{\m{i}\Gamma_N}{\pi}\m{Re}\left[\Psi\left(\frac{1}{2}+\m{i}\frac{\e-\delta_{S1}-\mu_N}{2\pi k_BT}\right)\right] \nonumber \\
&-\frac{\m{i}\Gamma_N}{\pi}\m{Re}\left[\Psi\left(\frac{1}{2}+\m{i}\frac{\e+U-\delta_{S1}-\mu_N}{2\pi k_BT}\right)\right] \nonumber \, .
\end{align}
For $W^{d d}_{0 0}(0,0)$, we get the same but $\delta_{S1}$ being replaced by $\delta_{S2}$.


\begin{thebibliography}{24}


\bibitem{josephson62}B. D. Josephson, Phys. Lett. {\bf 1}, 251 (1962).

\bibitem{tinkham96}M. Tinkham, \textit{Introduction to superconductivity}, 2nd edition, McGraw-Hill (1996).


\bibitem{buitelaar02}M.~R. Buitelaar, T. Nussbaumer, and C. Sch\"onenberger, Phys. Rev. Lett. {\bf 89}, 256801 (2002).

\bibitem{cleuziou06}J.-P. Cleuziou, W. Wernsdorfer, V. Bouchiat, T. Ondar\c{c}uhu, and M. Monthioux, Nat. Nanotech. {\bf 1}, 53 (2006).

\bibitem{eichler07} A. Eichler, M. Weiss, S. Oberholzer, C. Sch\"onenberger, A. Levy Yeyati, J.~C. Cuevas, and A. Mart\'{i}n-Rodero, Phys. Rev. Lett. {\bf 99}, 126602 (2007).

\bibitem{jorgensen07}  H.~I. J\o rgensen, T. Novotn\'y, K. Grove-Rasmussen,  K. Flensberg, and  P.~E. Lindelof, Nano Lett. {\bf 7}, 2441 (2007).

\bibitem{pillet10} J-D. Pillet, C.~H.~L. Quay, P. Morfin, C. Bena, A. Levy Yeyati, and P. Joyez, Nature Phys. {\bf 6}, 965 (2010). 

\bibitem{herrmann10} L.~G. Herrmann, F. Portier, P. Roche, A. Levy Yeyati, T. Kontos, and C. Strunk, Phys. Rev. Lett. {\bf 104}, 026801 (2010).

\bibitem{vandam06}J.~A van Dam, Yu.~V. Nazarov, E.~P.~A.~M. Bakkers, S. De Franceschi, and L.~P. Kouwenhoven, Nature {\bf 442}, 667 (2006).

\bibitem{sand07} T.~Sand-Jespersen, J. Paaske, B.~M. Andersen, K. Grove-Rasmussen, H.~I. J\o rgensen, M. Aagesen, C.~B. S\o rensen, P.~E. Lindelof, K. Flensberg, and J. Nyg\r{a}rd, Phys. Rev. Lett. {\bf 99}, 126603 (2007). 

\bibitem{hofstetter09}L. Hofstetter, S. Csonka, J.~Nyg\r{a}rd, and C. Sch\"onenberger, Nature {\bf 461}, 960 (2009).

\bibitem{heiblum12} A. Das, Y. Ronen, M. Heiblum, D. Mahalu, A.V. Kretinin, and H. Shtrikman, arXiv:1205.2455.

\bibitem{dirks11}T. Dirks, T. L. Hughes, S. Lal, B. Uchoa, Y-F. Chen, C. Chiavlo, P. M. Goldbart, and N. Mason, Nat. Phys. {\bf 7}, 386 (2011).


\bibitem{buizert07}C. Buizert, A. Oiwa, K. Shibata, K. Hirakawa, and S. Tarucha, Phys. Rev. Lett. {\bf 99}, 136806 (2007).


\bibitem{deacon10} R.~S. Deacon, Y. Tanaka, A. Oiwa, R. Sakano, K. Yoshida, K. Shibata, K. Hirakawa, and S. Tarucha, Phys. Rev. Lett. {\bf 104}, 076805 (2010).


\bibitem{kanai10} Y. Kanai, R.~S. Deacon, A. Oiwa, K. Yoshida, K. Shibata, K. Hirakawa, and S. Tarucha, Phys. Rev. B {\bf 82}, 054512 (2010). 


\bibitem{franceschi10}S. De Franceschi, L. Kouwenhoven, C. Sch\"onenberger, and W. Wernsdorfer, Nat. Nanotech. {\bf 1}, 703 (2010).

\bibitem{martin11}A. Mart\'{i}n-Rodero and A. Levy Yeyati, Adv. Phys. {\bf 60}, 899 (2011).


\bibitem{beenakker92}C. W. J. Beenakker and H. van Houten, \textit{Single-Electron Tunneling and Mesoscopic Devices}, edited by H. Koch and H. L\"ubbig (Springer, Berlin), pp. 175-179 (1992).


\bibitem{glazman89}L. I. Glazman and K. A. Matveev, Pis'ma Zh. Eksp. Teor. Fiz. {\bf 49}, 570 (1989) [JETP Lett. {\bf 49}, 659 (1989)].

\bibitem{rozhkov01} A. V. Rozhkov, D. P. Arovas, and F. Guinea, Phys. Rev. B {\bf 64}, 233301 (2001).



\bibitem{pala07} M. G. Pala, M. Governale, and J. K\"onig, New J. Phys. {\bf 9}, 278 (2007).

\bibitem{governale08}M. Governale, M.~G. Pala, and J. K\"onig, Phys. Rev. B {\bf 77}, 134513 (2008).



\bibitem{matsumoto01} D. Matsumoto, J. Phys. Soc. Jpn. {\bf 70}, 492 (2001).

\bibitem{vecino03} E. Vecino, A. Mart\'{i}n-Rodero and A. Levy Yeyati, Phys. Rev. B {\bf 68}, 035105 (2003).

\bibitem{rozhkov99}A. V. Rozhkov and D. P. Arovas, Phys. Rev. Lett. {\bf 82}, 2788 (1999).


\bibitem{siano04} F. Siano and R. Egger, Phys. Rev. Lett. {\bf 93}, 047002 (2004).

\bibitem{sellier05} G. Sellier, T. Kopp, J. Kroha, and Y. S. Barash, Phys. Rev. B {\bf 72}, 174502 (2005).

\bibitem{tanaka07} Y. Tanaka, A. Oguri, and A. C. Hewson, New J. Phys. {\bf 9}, 115 (2007).


\bibitem{karrasch08} C. Karrasch, A. Oguri, and V. Meden, Phys. Rev. B {\bf 77}, 024517 (2008).


\bibitem{meng09} T. Meng, S. Florens, and P. Simon, Phys. Rev. B {\bf 79}, 224521 (2009).






\bibitem{whan96}C. B. Whan and T. P. Orlando, Phys. Rev. B, {\bf 54}, 5255 (1996).

\bibitem{levy97} A. Levy Yeyati, J. C. Cuevas, A. L\'opez-D\'avalos, and A. Mart\'{i}n-Rodero, Phys. Rev. B {\bf 55}, 6137 (1997).

\bibitem{kang98} K.~Kang, Phys. Rev. B {\bf 57}, 11891 (1998).

\bibitem{avishai03}Y. Avishai, A. Golub, and A. D. Zaikin, Phys. Rev. B {\bf 67}, 041301 (2003).

\bibitem{dell08}L. Dell'Anna, A. Zazunov, and R. Egger, Phys. Rev. B {\bf 77}, 104525 (2008).


\bibitem{bratus95}E. N. Bratus, V. S. Shumeiko, and G. Wendin, Phys. Rev. Lett. {\bf 74}, 2110 (1995).

\bibitem{averin95} D. Averin and A. Bardas, Phys. Rev. Lett. {\bf 75}, 1831 (1995).

\bibitem{cuevas96} J.C. Cuevas, A. Martin-Rodero, and A. Levy Yeyati, Phys. Rev. B {\bf 54}, 7366 (1996).

\bibitem{sun02} Q.-F. Sun, H. Guo, and J. Wang, Phys. Rev. B {\bf 65}, 075315 (2002).

\bibitem{jonckheere09}T. Jonckheere, A. Zazunov, K. V. Bayandin, V. Shumeiko, and T. Martin, Phys. Rev. B {\bf 80}, 184510 (2009). 

\bibitem{perfetto09} E. Perfetto, G. Stefanucci, and M. Cini, Phys. Rev. B {\bf 80}, 205408 (2009).

\bibitem{stefanucci10} G. Stefanucci, E. Perfetto, and M. Cini, Phys. Rev. B {\bf 81}, 115446 (2010).

\bibitem{wu12}  B.H. Wu, J.C. Cao, and C. Timm, Phys. Rev. B {\bf 86}, 035406 (2012).

\bibitem{bruder94} C. Bruder and H. Schoeller, Phys. Rev. Lett. {\bf 72}, 1076 (1994).


\end{thebibliography}
\end{document}